\documentclass[showpacs,pra,twocolumn]{revtex4-1}
\usepackage[english]{babel}
\usepackage[utf8]{inputenc}
\usepackage{graphicx}
\usepackage{epstopdf}
\usepackage{footnote}
\usepackage{amsmath, amssymb}
\usepackage{dcolumn}
\usepackage{subfigure}
\usepackage{xcolor}

\begin{document}

\title{\bf Ramsey interferometry as a witness of acceleration radiation}

\author{Helder A. S. Costa$^{(a)}$ }%\email[]{} %hascosta@ufpi.edu.br
\author{Irismar G. da Paz$^{(a)}$}%\email[]{ }%irismarpaz@ufpi.edu.br
\author{Paulo R. S. Carvalho$^{(a)}$}%\email[]{ }%prscarvalho@ufpi.edu.br
\author{Marcos Sampaio$^{(b)}$}

\affiliation{(a) Universidade Federal do Piau\'{\i}, Departamento de
F\'{\i}sica, 64049-550 Teresina, PI, Brazil} \affiliation{(b)
Universidade Federal do ABC, Centro de Ciências Naturais e Humanas,
09210-580 Santo André, SP, Brazil. }

\begin{abstract}
We adapt a typical Ramsey interferometer  by inserting a linear
accelerator capable of accelerating an atom inside a single-mode
cavity. We demonstrate that this simple scheme allows us to estimate
the effects of acceleration radiation via interferometric
visibility. By using a Rydberg-like atom, our results suggest that,
for the transition regime of the order of GHz and interaction time
of 1 ns, acceleration radiation effects can be observable for
accelerations as low as $10^{17}\;\mathrm{m/s^2}$.

\end{abstract}

\maketitle
%%%%%%%%%%%%%%%%%%%%%%%%%%%%%%%%%%%%%%%%%%%%%%%%%%%%%%%
\section{Introduction}
  The Fulling-Davies-Unruh  effect \cite{Fulling} is the premise to demonstrate that
a linearly accelerated detector with acceleration $a$ in Minkowski
space on a hyperbolic (constant acceleration) trajectory responds to
the vacuum just as a detector at rest bathed by a Planck thermal
distribution of quanta of a field at Unruh temperature
$T_{\mathrm{U}} = \hbar a/(2\pi k_B c)$  \cite{Unruh} where $\hbar$
is the Planck's constant, $k_B$ the Boltzmann's constant and $c$ the
light speed in vacuum. On the other hand, it was shown
\cite{Scully,Belyanin} that virtual processes in which atoms jump to
an excited state while emitting a photon are an alternative way to
view Unruh acceleration radiation. In a few words, by breaking and
interrupting the virtual processes one can render the virtual
photons real. Moreover in \cite{PNAS}, it was shown that the
radiation emitted by atoms falling into a black hole resembles, but
is different from Hawking radiation, shedding light on the Einstein
principle of equivalence between acceleration and gravity. The
fundamental distinctions between Unruh effect and uniformly
accelerated atoms inside cavities are discussed in
\cite{ScullyReply}.

The process of atomic excitation by means of interaction with a
quantum field in vacuum can receive contributions from both vacuum
fluctuations and radiation reaction \cite{CT}. This, in turn, can be
extended to the study of  atoms under uniform acceleration. For
instance, the excitation rate and radiative energy shift of a
two-level atom in uniform acceleration coupled to massless scalar
field \cite{MSF} and electromagnetic field \cite{EF} in an inertial
frame were calculated in Minkowski vacuum and also in  spacetimes
with boundaries \cite{STB}. It was shown, for instance,  that
acceleration-induced perturbations lead to spontaneous excitation
even in vacuum. In \cite{ZHOU}, it was studied the  rate of change
of atomic energy for an atom in uniform acceleration coupled to
quantum electromagnetic field at a thermal state with temperature
$T$ from a co-accelerated observer viewpoint. They have shown that
the result  is the same of a local inertial observer  assuming the
Unruh temperature in the co-accelerated frame.

Because the Unruh temperature is smaller than $1\;\mathrm{K}$ even
for acceleration as high as $10^{21}$ $\mathrm{m/s^2}$, its direct
observation is not accessible with current technology. For this
reason, there has been growing interest in practical scheme or
experimental ideas to estimate the Unruh temperature \cite{Vanzella,
Martin-Martinez, Martin-Martinez2, Hu, Sabin}. Motivated by this
consideration, in this contribution we consider a typical single
particle Ramsey interferometer \cite{Ramsey}. We modify the
interferometer by inserting a linear accelerator (LA) (Cavity plus
short-pulse laser) capable of accelerating an atom through a
single-mode cavity. We show that this simple scheme allows us to
estimate the effects of acceleration radiation via fringe visibility
in a interferometric setup. In our theoretical analysis, we consider
a Rydberg-like atom initially prepared in the ground state. We find
that for the transition regime of the order of GHz and interaction
time of 1 ns, the visibility encodes information about the
acceleration radiation which  is manifested experimentally for
acceleration as low as $10^{17}$ $\mathrm{m/s^2}$. Since ultra-high
accelerations of Rydberg-like atoms have been obtained recently in
\cite{Eichmann, McWilliams}, the experimental implementation of the
present setup is attainable by current technology.

The outline of the paper is as follows. In Sec. 2, we start by
presenting a theoretical analysis of our setup and formulating an
effective model for the system which is a two-level atom interacting
simultaneously with a classical field and a quantum field. In Sec.
3, we evaluate finite-time corrections to the transition
probability. In Sec. 4, we show how our setup works and calculate
the interferometric visibility. The analysis and discussion of the
results are presented in Sec. 5. Finally, our concluding remarks are
addressed in Sec. 6.

%%%%%%%%%%%%%%%%%%%%%%%%%%%%%%%%%%%%%%%%%%%%%%%%%%%%%%%%

 \section{The setup}
\begin{figure}[h]
\centering
\includegraphics[height=3.5cm]{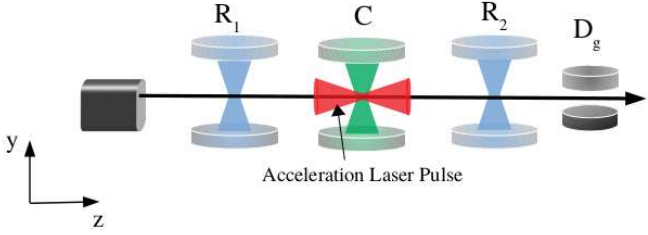}
\caption{The proposed scheme for single-particle interference acting
as a witness of acceleration radiation. A key requirement in the practical realization of the acceleration mechanism is the ability to produce a short-pulse laser with intensity and frequency separately controlled \cite{Eichmann, McWilliams}. In addition, the short-pulse laser must be of low enough intensity so that it does not ionize or strongly perturb the internal states of the atom.}\label{fig1}
\end{figure}

 Consider a conventional Ramsey interferometer
\cite{Ramsey} for accelerated two-level atoms. Suppose that such
two-level atoms are produced by the source in the ground $|g\rangle$
or excited $|e\rangle$ states and enter in the first Ramsey zone
(cavity $R_1$) which has a field resonant or quasi-resonant with the
transition $|g\rangle \Leftrightarrow |e\rangle$ yielding a $\pi/2$
pulse on the atoms \cite{Ramsey,Haroche}. After the atoms pass
through the cavity $R_1$, they enter into  cavity $C$ where they are resonantly coupled with a single mode electromagnetic field. Inside cavity $C$, the atoms are linearly accelerated by a short-pulse laser. After that, the atoms enter in the second Ramsey zone (cavity $R_2$) where their internal states
are recombined and the interference pattern measured in the detector
$D_g$. This setup is sketched in Fig. \ref{fig1}.

The Hamiltonian of the system includes two interaction terms. One of
them is the quantum interaction of the atom and the quantum field
inside the cavity C and the other one is the semi-classical
interaction of the atom and the classical field produced by the
laser. The total Hamiltonian is described by the usual
Jaynes-Cummings Hamiltonian \cite{JC}
\begin{align}
\hat{H} = \hat{H}_a + \hat{H}_f +
\hat{H}_{\mathrm{int}}^{\mathrm{Q}} + \hat{H}_{\mathrm{int}}^{\mathrm{C}},
\end{align}
where $\hat{H}_a = \hbar\omega\hat{\sigma}_z$ is the Hamiltonian for a two-level atom with
upper  and lower levels $|e\rangle$ and $|g\rangle$, respectively, with $\hat{\sigma}_z = \frac{1}{2}[|g\rangle\langle g| - |e\rangle\langle e|]$. $\hat{H}_f = \sum_k \hbar\nu_k\hat{a}^{\dagger}_k\hat{a}_k$ is the
Hamiltonian of the field where $\hat{a}_k^{\dagger}$ and $\hat{a}_k$ are
photon creation and annihilation operators, and
$\hat{H}_{\mathrm{int}}^{\mathrm{Q}}$ stands for the coupling of the
atom with the quantum field inside the cavity C which in the atomic frame of reference reads
\begin{align} \label{Hint}
\hat{H}_{\mathrm{int}}^{\mathrm{Q}} = \hbar\sum_k \tilde{\lambda}_k(\tau)
[\hat{a}_k^{\dagger}\hat{\sigma}_{-}e^{i\nu_k t(\tau) - ikz(\tau) -
i\omega\tau} + \mathrm{h}. \mathrm{c}.].
\end{align}
In equation above, $\tau$ is the atom proper time ($t$ denotes the time in the
inertial laboratory frame) and $\tilde{\lambda}_k(\tau)$ is the
effective coupling constant between the atom and the field, which
depends on the dipole moment and the field amplitude. Notice that
the Hamiltonian (\ref{Hint}) describes the atom-field interaction in
the dipole and rotating-wave approximations \cite{JC}. We also take
$\tilde{\lambda}_k (\tau) = \lambda_k W (\tau)$, where we choose $W
(\tau) = \exp\left(-\frac{|\tau|}{T}\right)$ as a gradual window
function \cite{Higuchi, Sriramkumar} in order to describe the
interaction between atom and field for a finite time interval during
the flight in the linear accelerator. $T$ is a characteristic time
such that $W(\tau \ll T) \approx 1$  and $W(\tau \gg T) \approx 0$.
On the other hand, $\hat{H}_{\mathrm{int}}^{\mathrm{C}}$ stands for
the coupling between the atom and the classical laser beam which in
the dipole approximation is given by
\begin{align}
\hat{H}_{\mathrm{int}}^{\mathrm{C}} = \hbar\kappa\left(
e^{-i\omega_{\mathrm{L}}t}\hat{\sigma}_{+} + e^{i\omega_{\mathrm{L}}t}\hat{\sigma}_{-}\right),
\end{align}
where $\omega_{\mathrm{L}}$ is the laser beam frequency and $\kappa$
the coupling constant. $\hat{\sigma}_{+} = |e\rangle\langle g|$ and $\hat{\sigma}_{-} = |g\rangle\langle e|$ are the atomic raising and lowering operators, respectively. 

\begin{figure}[h]
\centering
\includegraphics[height=3.5cm]{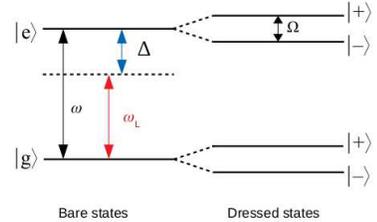}
\caption{Schematic representation of the dipole transition $|g\rangle \rightarrow |e\rangle$ at frequency $\omega$ coupled to the laser at frequency $\omega_{\mathrm{L}}$ and the semiclassical dressed states.  }\label{States}
\end{figure}

 The total Hamiltonian of the system in the reference frame rotating at the frequency $\omega_{\mathrm{L}}$ of the classical field, has the form
 \begin{align*}
 \hat{H}^{\mathrm{RF}} &= \hat{\mathcal{H}}_0 - \hbar\sum_k \delta_k\hat{a}^{\dagger}_k\hat{a}_k \\
 &+ \hbar\sum_k \tilde{\lambda}_k(\tau)[\hat{a}_k^{\dagger}\hat{\sigma}_{-}e^{i\nu_k t(\tau) - ikz(\tau) -
i\omega\tau} + \mathrm{h}. \mathrm{c}.],
 \end{align*}
 where $\delta_k = \omega_{\mathrm{L}} - \nu_k$, and $\hat{\mathcal{H}}_0 = -\hbar\Delta\hat{\sigma}_z + \hbar\kappa(\hat{\sigma}_{+} + \hat{\sigma}_{-})$ with detuning $\Delta = \omega_{\mathrm{L}} - \omega$. The presence of the laser change the atom states $|e\rangle$ and $|g\rangle$ to the dressed states $|+\rangle$ and $|-\rangle$, respectively, see Fig. \ref{States}. 
 
 Let the coupling to the laser be much stronger than the coupling to the cavity modes, i.e., $|\kappa| \gg \lambda_k$. Thus, we can express $\hat{H}^{\mathrm{RF}}$ in the basis of eigenstates $|\pm \rangle$ of $\hat{\mathcal{H}}_0$ \cite{Pielawa}, with 
\begin{align*}
 \hat{\mathcal{H}}_0 |\pm \rangle = -\frac{\hbar(\Delta \mp \Omega)}{2}|\pm \rangle 
\end{align*} 
 where $\Omega = \sqrt{\Delta^2 + 4\kappa^2}$ is called the effective Rabi frequency. In the limit where the detuning $\Delta$ is large as compared to the $|\kappa|$, the effective Rabi frequency is given by $\Omega \approx \Delta + \delta\omega$ where $\delta\omega = \frac{2\kappa^2}{\Delta}$ corresponds to the shift in the transition frequency due to the interaction of the atom with the laser.
 
 The semiclassical dressed states are given by
 \begin{align*}
 |+\rangle &= \sin\theta|g\rangle + \cos\theta|e\rangle, \\
 |-\rangle &= \cos\theta|g\rangle - \sin\theta|e\rangle,
\end{align*}  
 where $\cos\theta = \sqrt{\frac{\delta\omega}{2(\Delta + \delta\omega)}}$ and $\sin\theta = \sqrt{\frac{2\Delta + \delta\omega}{2(\Delta + \delta\omega)}}$.  Let's introduce the raising and lowering operators for the dressed states basis by $\hat{\pi}_{+} = |+\rangle\langle -|$ and $\hat{\pi}_{-} = |-\rangle\langle +|$. By using the semiclassical dressed state basis and operator notation we can rewrite $\hat{H}^{\mathrm{RF}}$ in the interaction picture as
 \begin{align*}
 \hat{H}^{I} &= \hbar \sum_k\tilde{\lambda}_k(\tau)\{\sin\theta\cos\theta[\hat{a}_k(\tau)e^{-i\delta_kt} + \hat{a}_k^{\dagger}(\tau)e^{i\delta_kt}]\hat{\pi}_{z} \\
 &+ \cos^2\theta[\hat{a}_k(\tau)\hat{\pi}_{+}e^{i(\Omega - \delta_k)t} + \hat{a}_k^{\dagger}(\tau)\hat{\pi}_{-}e^{-i(\Omega - \delta_k)t}] \\
 &- \sin^2\theta[\hat{a}_k^{\dagger}(\tau)\hat{\pi}_{+}e^{i(\Omega + \delta_k)t} + \hat{a}_k(\tau)\hat{\pi}_{-}e^{-i(\Omega + \delta_k)t}]\},
\end{align*}  
 where $\hat{a}_k(\tau) = \hat{a}_ke^{-i\nu_k t(\tau) + ikz(\tau) + i\omega\tau}$ and $\hat{a}_k^{\dagger}(\tau) = \hat{a}_k^{\dagger}e^{i\nu_k t(\tau) - ikz(\tau) - i\omega\tau}$. For $\delta_k = \Omega$ and if $|\lambda_k| \ll \Omega$ (strong driven regime), we can realize a rotating-wave approximation and choose from $\hat{H}^{I}$ the resonant terms
 \begin{align} \label{Heff}
 \hat{H}_{\mathrm{eff}} = \hbar \sum_k\tilde{\lambda}_k(\tau)\cos^2\theta[\hat{a}_k(\tau)\hat{\pi}_{+} + \hat{a}_k^{\dagger}(\tau)\hat{\pi}_{-}]
\end{align}

%In addition, the laser beam causes a light shifted (Stark shift) in
%the atomic levels creating an additional potential for the particle
%which produces a dipole force that accelerates the atom

In the limit of large detuning $\Delta \gg |\kappa|$ the atom is uniformly accelerated due a dipole force produced by the laser pulse \cite{Sasura, Obada}.  For the uniformly accelerated atom traveling in the $z$-direction,
the relation between the laboratory Minkowskian inertial frame  with
the accelerated frame is given by
\begin{eqnarray}
 t(\tau) = \frac{c}{a}\sinh(\frac{a\tau}{c}), \quad z(\tau) =
\frac{c^2}{a}\cosh(\frac{a\tau}{c}), \nonumber
\end{eqnarray}
where $a$ is the acceleration. Thus the uniformly accelerated atom
describes hyperbolic trajectories in the right-hand half Minkowski
spacetime known as the right Rindler wedge defined by $z > |t|$
(see, e.g., Ref. \cite{Rindler}). Moreover, for copropagating atom
and field, $k_z = k = \nu_k/c$ and $\lambda_k$ scales as  $
\lambda_k = \lambda_k e^{-\frac{a\tau}{c}}$ for a uniformly
accelerated atom due to the eletric field transformation properties
to the atomic frame \cite{Scully, Belyanin}. The exponent in
(\ref{Hint}) becomes
\begin{eqnarray}
 i\nu_k t(\tau) - ikz(\tau) &=& \frac{i\nu_k c}{a}\sinh(\frac{a\tau}{c}) - \frac{ikc^2}{a}\cosh(\frac{a\tau}{c}) \nonumber \\
 &=& -\frac{i\nu_k c}{a}e^{-\frac{a\tau}{c}}. \nonumber
\end{eqnarray}
Therefore, the effective interaction Hamiltonian (\ref{Heff}) can be rewritten as
\begin{align} \label{Hint2}
\hat{H}_{\mathrm{eff}} = \hslash \sum_k \tilde{\lambda}_k\cos^2\theta
\Big[\hat{a}_k^{\dagger}\hat{\pi}_{-} e^{ -\frac{i\nu_k
c}{a}e^{-\frac{a\tau}{c}} - i\omega\tau -\frac{a\tau}{c}} +
\mathrm{h}. \mathrm{c}.\Big],
\end{align}
 For simplicity, we assume the single mode approximation
for the coupling between the atom and the quantum field in the
cavity C as in conventional derivations of the Jaynes-Cummings
model \cite{Scully2}. Thus we may ignore the coupling between the
atom and all field modes except with the mode $k$ with frequency
equal to $\nu_k$. It is worth mention that this approximation produces
accurate results if the evolution times are long (for details see,
e.g., Ref. \cite{Brown}).

%%%%%%%%%%%%%%%%%%%%%%%%%%%%%%%%%%%%%%%%%%%%%%%%%%%%%%%%%%

\section{Leading order correction to the transition amplitude}
In the interaction picture, the leading order unitary transformation
induced by the Hamiltonian (\ref{Hint2}) in the weak coupling regime reads
\begin{align}\label{unitary}
\hat{U} &\approx 1 - iI(a)\cos^2\theta\hat{a}_k^{\dagger}\hat{\pi}_{-} - iI^{*}(a)\cos^2\theta\hat{a}_k\hat{\pi}_{+},
\end{align}
where the transition amplitude is
\begin{eqnarray}
I(a) = \lambda_{k} \int_{-\infty}^{\infty}d\tau \exp\Big[-\frac{i\nu_{k}
c}{a}e^{-\frac{a\tau}{c}} - i\omega\tau -\frac{a\tau}{c} -
\frac{|\tau|}{T}\Big]. \nonumber
\end{eqnarray}
 By making the substituiton $x = \frac{\nu_{k} c}{a}\,e^{-\frac{a\tau}{c}}$, $I(a)$ reduces to
\begin{eqnarray}
I(a) &=& \frac{\lambda_{k}}{\nu_{k}}\left(\frac{a}{\nu_{k}
c}\right)^{\frac{i\omega c}{a} -
\frac{c}{aT}}\int^{\infty}_{\frac{\nu_{k} c}{a}}dx\; x^{\frac{i\omega
c}{a} - \frac{c}{aT}} \; e^{-ix} \nonumber \\ &+&
\frac{\lambda_{k}}{\nu_{k}}\left(\frac{a}{\nu_{k} c}\right)^{\frac{i\omega c}{a}
+ \frac{c}{aT}}\int_{0}^{\frac{\nu_{k} c}{a}}dx \; x^{\frac{i\omega
c}{a} + \frac{c}{aT}} \; e^{-ix}.
\end{eqnarray}
Recalling the definition of incomplete gamma functions\footnote{The upper and lower incomplete gamma functions are defined, respectively, by $$\Gamma(s,x) = \int_{x}^{\infty} t^{s-1}e^{-t}dt,$$ $$\gamma(s,x) = \int_{0}^{x} t^{s-1}e^{-t}dt.$$}
\cite{Zubair1994} enables us to write $I(a)$ as
\begin{eqnarray} \label{I3}
I(a) &=& -\frac{i\lambda_{k}}{\nu_{k}}\left(\frac{a}{\nu_{k}
c}\right)^{\frac{i\omega c}{a}}e^{-\frac{\pi \omega
c}{2a}}\Bigg[\left(\frac{a}{\nu_{k}
c}\right)^{-\frac{c}{aT}}e^{\frac{i\pi c}{2aT}} \Gamma\Big(1 +
\mu_{-}, \frac{\nu_{k} c}{a}\Big) \nonumber \\ &+& \big(\frac{a}{\nu_{k}
c}\big)^{\frac{c}{aT}}e^{-\frac{i\pi c}{2aT}} \gamma\Big(1 +
\mu_{+}, \frac{\nu_{k} c}{a}\Big)\Bigg].
\end{eqnarray}
where $\mu_{\pm} = \frac{i\omega c}{a} \pm \frac{c}{aT} $. The
expression above can be studied using the well known properties and
asymptotic  limits of incomplete gamma functions. In particular, let
us consider the limit of high acceleration in which one obtains
\begin{eqnarray}
\Gamma(1 + \mu_{-}, \frac{\nu_{k} c}{a} \rightarrow 0) &=& \Gamma(1 + \mu_{-}), \nonumber \\
\gamma(1 + \mu_{+}, \frac{\nu_{k} c}{a} \rightarrow 0) &=& 0. \nonumber
\end{eqnarray}
Now, substituting these results in (\ref{I3}), yields the transition
probability:
\begin{eqnarray}
|I(a)|^2_{\;\;\text{large $a$}} =
\frac{\lambda_{k}^2}{\nu_{k}^2}\left(\frac{a}{\nu_{k}
c}\right)^{-\frac{2c}{aT}}e^{-\frac{\pi \omega c}{a}} \Gamma(1 +
\mu_{-})\Gamma(1 + \mu_{-}^{*}). \nonumber
\end{eqnarray}
Furthermore we can expand in terms of ($\frac{1}{a T}$) in order to
evaluate finite-time corrections to the transition probability.
Thus, for $\frac{c}{aT} \ll 1$  and by using the relation  $\Gamma(1
- \frac{i\omega c}{a})\Gamma(1 + \frac{i\omega c}{a}) =
\frac{\pi\omega
    c}{a}\sinh^{-1}(\frac{\pi\omega c}{a})$, the transition probability $|I(a)|^2$ can be written as
\begin{eqnarray}
|I(a)|^2 &\approx & \frac{2\pi \lambda_{k}^2\omega
c}{a\nu_{k}^2}\frac{1}{e^{\frac{2\pi \omega c}{a}} - 1}\Bigg\{ 1 -
\frac{c}{aT}\Bigg[2\ln\left(\frac{a}{\nu_{k} c}\right) \nonumber \\ & +&
\psi\left(\Omega_{+}\right)  + \psi\left(\Omega_{-}\right)\Bigg]  +
{\mathcal{O}}(aT)^{-2}\Bigg\}, \label{probtrans}
\end{eqnarray}
where $\psi (z)$ is the digamma function and $\Omega_{\pm} = 1 \pm
\frac{i\omega c}{a}$. It is clear that in the $\frac{c}{aT}
\rightarrow 0$ limit, we recover the thermal spectrum as in the
Unruh effect in free space. In other words, the radiation emitted by
the accelerated atom in the cavity C corresponds to thermal
radiation with temperature equal to the Unruh temperature
$T_{\mathrm{U}}$.

%%%%%%%%%%%%%%%%%%%%%%%%%%%%%%%%%%%%%%%%%%%%%%%%%%%%%

\section{Ramsey Interferometry}
In order to explain how the apparatus depicted in  Fig. \ref{fig1}
works, let us assume that initially the system (atom plus quantum
filed in cavity C) is prepared in the state $|\psi_1\rangle =
|0_{\mathcal{M}}\rangle\otimes|g\rangle$, where
$|0_{\mathcal{M}}\rangle$ is the Minkowski vacuum state of the field
defined in the inertial laboratory frame. The interaction between
the atom and the cavity $\mathrm{R}_1$ is chosen to produce a
$\mathrm{\pi}/2$ rotation in the Bloch sphere. This results in a
transformation of the $|g\rangle$ state to the superposition state:
\begin{eqnarray}
|\psi_1\rangle \rightarrow
\frac{1}{\sqrt{2}}|0_{\mathcal{M}}\rangle\otimes[|g\rangle +
|e\rangle]. \nonumber
\end{eqnarray}
The probability amplitude of finding the atom in $|g\rangle$ or
$|e\rangle$ between the cavities $\mathrm{R}_1$ and $\mathrm{R}_2$
accumulates a quantum phase. Thus, a phase shift $\Phi$ is
introduced to the atomic state if the atom is in $|e\rangle$:
\begin{eqnarray}\label{psi2}
|\psi_2\rangle \rightarrow
\frac{1}{\sqrt{2}}|0_{\mathcal{M}}\rangle\otimes[|g\rangle +
e^{i\Phi}|e\rangle]. 
\end{eqnarray}
Now, let us consider that the two-level atoms cross a microwave cavity (cavity C) where they are linearly accelerated in the $z$-direction by a short-pulse laser, as shown in Fig. \ref{fig1}. From (\ref{unitary}), we find that the interaction between the uniformly accelerated atom and the field mode $k$ produces the following transformations
\begin{align}
|0_{\mathcal{M}}\rangle\otimes|-\rangle &\rightarrow |0_{\mathcal{M}}\rangle\otimes|-\rangle,\nonumber\\
|0_{\mathcal{M}}\rangle\otimes|+\rangle &\rightarrow |0_{\mathcal{M}}\rangle\otimes|+\rangle -
iI(a)\cos^2\theta|1_{\mathcal{M}}\rangle\otimes|-\rangle. \label{atomfield}
\end{align}
Such transformations  show that when the uniformly accelerated atom is in the ground-state and the LA in the vacuum state, the coupling with the field mode in the LA does not affect the atom-field state. On the other hand, when the atom is in the excited state,  a photon can be emitted inside the LA in the vacuum state which is called acceleration radiation \cite{Scully}. This mimics the superposition of an inertial and accelerated trajectory in a interferometer as discussed in \cite{Martin-Martinez}.
Since we are interested in observing only the effect of the
acceleration radiation in the interference pattern, we consider that
the transition probability from the excited to the ground state or
from the ground to the excited state is negligible when the
non-accelerated atom interacts with the cavity $C$ in the vacuum
state, i.e., when the short-pulse laser is off. This is consistent
with an atom-field interaction in the weak coupling regime and for
very short interaction times \cite{Schleichbook}. By writing (\ref{psi2}) in the $\{|+\rangle, |-\rangle\}$ basis and using Eq. (\ref{atomfield}), after the atom passes through the cavity
C and propagates to cavity $R_2$, the state of the system reads 
\begin{align}
|\psi_3\rangle &\rightarrow
\frac{1}{\sqrt{2}}[|0_{\mathcal{M}}\rangle\otimes(|g\rangle + e^{i\Phi}|e\rangle) \nonumber \\
&- iI(a)(e^{i\Phi}\cos\theta + \sin\theta)\cos^3\theta|1_{\mathcal{M}}\rangle\otimes|g\rangle \nonumber \\
&+ iI(a)(e^{i\Phi}\cos\theta + \sin\theta)\cos^2\theta\sin\theta|1_{\mathcal{M}}\rangle\otimes|e\rangle]. \nonumber
\end{align}
 Just as for the cavity $\mathrm{R}_1$, the interaction time between the atom and the cavity $\mathrm{R}_2$ is
chosen so that it corresponds to a $\mathrm{\pi}/2$-pulse
interaction. Finally, after the atom passes through the cavity
$\mathrm{R}_2$, the final state of the system becomes
\begin{align}
|\psi_4\rangle &\rightarrow e^{i\frac{\Phi}{2}}\Bigg\{ |0\rangle\otimes\left[\cos\left(\frac{\Phi}{2}\right)|g\rangle + i\sin\left(\frac{\Phi}{2}\right)|e\rangle\right]  \nonumber \\
& -iI(a)\mathcal{F}_{-}|1\rangle\otimes|g\rangle - iI(a)\mathcal{F}_{+}|1\rangle\otimes|e\rangle \Bigg\}, \label{Psi4}
\end{align}
 where 
  \begin{align*}
  \mathcal{F}_{\pm}(\theta, \Phi) = \frac{1}{2}\cos^2\theta(e^{\frac{i\Phi}{2}}\cos\theta + e^{-\frac{i\Phi}{2}}\sin\theta)(\cos\theta \pm \sin\theta).
  \end{align*}
  The reduced density matrix $\hat{\rho}_{\mathrm{A}} =
\mathrm{Tr}_{\mathcal{M}}[|\psi_4\rangle\langle\psi_4|]$ of the atom
is obtained using (\ref{Psi4}) and tracing out the field degrees of
freedom. After some algebraic manipulations one obtains
\begin{align}
\hat{\rho}_{\mathrm{A}} &=  \frac{1}{N}\Bigg\{\left[|I(a)|^2|\mathcal{F}_{-}|^2 + \cos^2\left(\frac{\Phi}{2}\right)\right]|g\rangle\langle g| \nonumber \\
&+ \left[|I(a)|^2\mathcal{F}_{+}\mathcal{F}_{-}^{*} + i\sin\left(\frac{\Phi}{2}\right)\cos\left(\frac{\Phi}{2}\right)\right]|e\rangle\langle g| \nonumber \\
&+ \left[|I(a)|^2\mathcal{F}_{+}^{*}\mathcal{F}_{-} - i\sin\left(\frac{\Phi}{2}\right)\cos\left(\frac{\Phi}{2}\right)\right]|g\rangle\langle e| \nonumber \\
&+ \left[|I(a)|^2|\mathcal{F}_{+}|^2 + \sin^2\left(\frac{\Phi}{2}\right)\right]|e\rangle\langle
e| \Bigg\},\nonumber
\end{align}
 where the normalization factor $N$ is given by
\begin{align*}
N = 1 + |I(a)|^2|\mathcal{F}_{+}|^2 +  |I(a)|^2|\mathcal{F}_{-}|^2,
\end{align*}
and $|I(a)|^2$ is given by (\ref{probtrans}). 

From $\hat{\rho}_{\mathrm{A}}$, we find that the probability
to detect each atom in the $|g\rangle$ state by the detector $D_{g}$
is given by 
\begin{align}
P_g &= \frac{1}{N}\Bigg\lbrace \cos^2\left(\frac{\Phi}{2}\right) + \frac{|I(a)|^2}{4}\cos^4\theta[1 + \sin 2\theta(\cos\Phi - 1) \nonumber \\
& + \cos\Phi\sin^22\theta]\Bigg\rbrace , \nonumber \\
& = \frac{1}{N}\Bigg\lbrace \cos^2\left(\frac{\Phi}{2}\right) + \frac{|I(a)|^2}{4}\left(\frac{\delta\omega}{2(\Delta + \delta\omega)}\right)^2 \nonumber \\
& \times \left[1 + \frac{2\kappa}{(\Delta + \delta\omega)}(\cos\Phi - 1) 
 + \frac{4\kappa^2}{(\Delta + \delta\omega)^2}\cos\Phi\right] \Bigg\rbrace,
\end{align}
which is the Ramsey interference for the accelerated two-level
atoms. The fringe visibility for the interference pattern is defined
by $V = \frac{P_g^{\mathrm{max}} -
P_g^{\mathrm{min}}}{P_g^{\mathrm{max}} + P_g^{\mathrm{min}}}$, where
the max/min values are calculated with respect to phase $\Phi$. The
probability $P_g$ has the maximum (minimum) value when $\Phi = 0$
($\Phi = \pi$). A straightforward calculation shows that the
visibility is given by
\begin{eqnarray}\label{vis} 
 V &=& \frac{1 + |I(a)|^2\left(\frac{\delta\omega}{2(\Delta + \delta\omega)}\right)^2 \frac{2\kappa}{\Delta + \delta\omega} \left(1 - \frac{2\kappa}{\Delta + \delta\omega}\right)}{1 + |I(a)|^2\left(\frac{\delta\omega}{2(\Delta + \delta\omega)}\right)^2 \left(1 - \frac{2\kappa}{\Delta + \delta\omega}\right) }.
\end{eqnarray}
Our first observation is that the intensity and the corresponding
fringe visibility measured by detector $D_g$ provide us a
straightforward way of estimating the effects of acceleration by
means of interference properties. It is known that de-excitation
(excitation) of the atom by photon emission (absorption) is a source
of loss of coherence \cite{Viale}. Here, we can observe from Eq.
(\ref{vis}) that, because of the acceleration radiation emitted by
the accelerated two-level atoms, loss of coherence effects are
apparent showing a reduction in the fringe visibility, i.e, $V<1$
(partially coherent). In the limit $a \rightarrow 0$ and $\kappa
\rightarrow 0$ (short-pulse laser off), we get the visibility equal
to $1$ (completely coherent). We have to emphasize that this source of loss of coherence is produced by the dipole interaction which becomes acceleration dependent as a consequence of the coordinates transformations relating the inertial and accelerated frames. 

%%%%%%%%%%%%%%%%%%%%%%%%%%%%%%%%%%%%%%%%%%%%%%%%%%%%%%%%%

\section{Results and discussion}
The main  point of interest in our discussion is to show how the
interaction between an uniformly accelerated two-level atom and an
electromagnetic field in the vacuum state affects interference
properties. From a practical point of view, let us consider
two-level atoms such as Rydberg atoms for which the transition
frequency between the levels is of the order of GHz \cite{Raimond}. In addition,
we consider timescale of evolution much smaller than the lifetime of
Rydberg states. To verify the loss of coherence effects induced by
acceleration radiation, we compare the fringe visibility when the
atom is inertial (short-pulse laser off) with the visibility when
the atom is accelerated (short-pulse laser on). We define the
visibility difference between the two situations as $\delta V = V(0)
- V(a)$.

\begin{figure}[h]
\centering
\includegraphics[height=3.5cm]{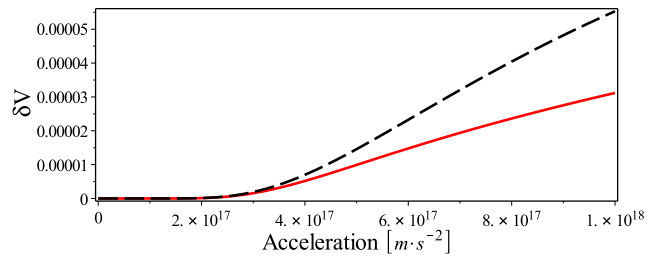}
\caption{Visibility difference as a function of acceleration for 
 $T = 1\;\mathrm{ns}$ red (solid) line and $T \rightarrow \infty$ black (dashed) line. Here we have fixed $\lambda_k = 50$ MHz, $\kappa = 200$ MHz, $\omega_{\mathrm{L}} = 0.5$ GHz, $\delta\omega = 160$ MHz and $\omega = \nu_k = 1$ GHz.} \label{fig3}
\end{figure}

In Fig. \ref{fig3}, we plot the visibility difference $\delta V$ as
a function of the acceleration for two different interaction
times $T = 1$ ns red (solid) line and $T \rightarrow \infty$ black (dashed) line.
We consider the microwave regime of cavity field, i.e., $\omega$ and $\nu_k$ of the order of GHz, 
and for the coupling constants $\lambda_k = 50$ MHz and $\kappa = 200$ MHz,  where
$\lambda_k \ll \omega$ (weak coupling regime) to guarantee the validity of the approximations used in Eq. (\ref{unitary}).

We can observe that the difference between the fringe
visibility of the accelerated and no accelerated interference
pattern, monotonically increases when the acceleration grows. 
The loss in the fringe visibility can be
understood as a result of the change in the internal state induced
by acceleration radiation emitted due to the dipole approximation. In particular, the Fig. \ref{fig3}
shows that the visibility difference approaches to the
value $\delta V \approx 10^{-5}$ for acceleration of the order of
$5\times10^{17}\;\mathrm{m\cdot s^{-2}}$. Notice that, for an
acceleration of $\approx 5\times10^{17}\;\mathrm{m\cdot s^{-2}}$ and an
interaction time of the order of $1\;\mathrm {ns}$, the cavity length is approximately $\approx 25\;\mathrm{cm}$. The black dashed line in Fig. \ref{fig3} shows the visibility
difference as a function of acceleration for a long interaction time ($T \rightarrow \infty$). This asymptotic regime can be considered as an ideal curve. However, it is noteworthy that in the high acceleration regime the increase in interaction time corresponds to an increase in cavity size and most relativistic the atom becomes.

\begin{figure}[h]
\centering
\includegraphics[height=3.5cm]{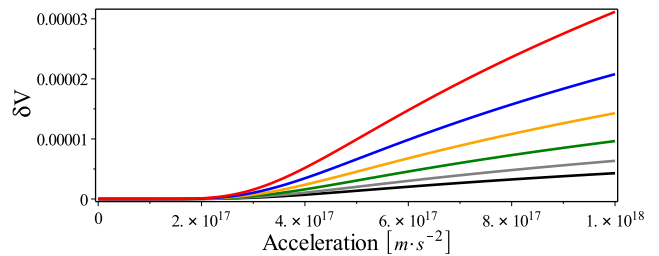}
\caption{Visibility difference as a function of acceleration for
different values of the $\kappa$ and of the $\delta\omega$ . (black line) $\kappa = 150$ MHz and $\delta\omega = 90$ MHz, (gray line) $\kappa = 160$ MHz and $\delta\omega = 102$ MHz, (green line) $\kappa = 170$ MHz and $\delta\omega = 116$ MHz, (orange line) $\kappa = 180$ MHz and $\delta\omega = 130$ MHz, (blue line) $\kappa = 190$ MHz and $\delta\omega = 144$ MHz, and (red line) $\kappa = 200$ MHz and $\delta\omega = 160$ MHz.  Here we have fixed $\lambda_k = 50$ MHz, $\omega_{\mathrm{L}} = 0.5$ GHz, and $\omega = \nu_k = 1$ GHz.}\label{fig4}
\end{figure}

 In Fig. \ref{fig4} we plot the visibility difference as a function of acceleration for different
values of $\kappa$ and consequently of $\delta\omega$ in order to observe the influence of the shift in the transition frequency on the visibility difference. We can observe that the visibility difference increases when $\kappa$ and $\delta\omega$ increase.  
These results show that although the coupling with the classical laser pulse does not produce entanglement and decoherence the atom states get dressed producing a shift in the transition frequency which contributes to the loss of coherence. 

Therefore, the proposed experimental setup which is based in a type
of Ramsey interferometry with accelerated two-level atoms can be
used to access acceleration radiation effects by measuring the
reduction in the fringe visibility. In addition, we would like to
emphasize that our setup is idealized, since we have considered
cavities with perfect mirrors. Investigation of the imperfection
effects will be leaved for a future work.

\section{Conclusions}
In Ref. \cite{Martin-Martinez}, it  was proposed the detection of
the Unruh effect by measuring the difference in the Berry's phase of
accelerated and inertial detectors. It was found that the Unruh
effect could be detected for accelerations as low as
$10^{17}\;\mathrm{m/s^2}$. Although that proposal could be tested in
any experimental setup capable to measure the Berry's phase a
specific implementation was left for a future work. Here, we present
a specific implementation to detect the analogous of this effect by
measuring the fringe visibility of the Ramsey interferometry for
accelerated two level atoms instead of Berry's phase. We show that
acceleration radiation emitted in a cavity in the vacuum state
produces loss of coherence which can be observed by measuring the
reduction of the fringe visibility. It is also shown that our setup
can in principle detect a difference of $10^{-5}$ in the fringe
visibility of Rydberg atoms with transition frequency of the order
of GHz and accelerations as low as
$5\times10^{17}\;\mathrm{m/s^2}$.

\section*{Acknowledgments}
HASC would like to thank the Brazilian funding agency CAPES for
financial support. PRSC would like to thank CNPq (Brazilian funding
agency) through grant Universal-431727/2018. MS acknowledges CNPq
for a research grant.

\end{document}